\documentclass[twocolumn,prd]{revtex4}

\usepackage{graphicx}
\usepackage{dcolumn}
\usepackage{bm}

\newcommand{\be}{\begin{equation}}
\newcommand{\ee}{\end{equation}}
\newcommand{\ba}{\begin{eqnarray}}
\newcommand{\ea}{\end{eqnarray}}

\begin{document}
\title{Large Scale  Magnetic Fields and the Number of Cosmic Ray Sources
above $10^{19}$ eV}
\author{Claudia Isola$^{a,b}$, G{\"u}nter Sigl$^b$}
\affiliation{$^a$ Centre de Physique Th{\'e}orique,
Ecole Polytechnique, 91128 Palaiseau Cedex, France\\
$^b$ GReCO, Institut d'Astrophysique de Paris, C.N.R.S., 98 bis boulevard
Arago, F-75014 Paris, France}

\date{March 14, 2002}

\begin{abstract}
We present numerical simulations for the two-point
correlation function and the angular power spectrum of nucleons
above $10^{19}$~eV injected by a discrete distribution of sources
following a simple approximation to the profile of the Local Supercluster.
We develop a method to constrain the number of sources
necessary to reproduce the observed sky distribution of ultra-high energy
cosmic rays, as a function of the strength of the large scale cosmic
magnetic fields in the Local Supercluster. While for fields
$B\lesssim0.05 \mu\,$G the Supercluster source distribution appears
inconsistent with the data for any number of sources,
fields of strength $B\simeq0.3 \mu\,$G could reproduce the observed data
with a number of sources around 10.
\end{abstract}

\maketitle

\section{Introduction}
Despite a growing amount of data the origin of cosmic rays especially
at the highest energies is still obscure. For ultra-high energy cosmic
rays (UHECR) with energies above $10^{18}$,
there are still many open questions such as ``How can particles
be accelerated to these extremely high energies?'',
and ``What are their sources~?''~\cite{reviews}. The best candidates
for acceleration sources are powerful objects, such as
hot spots of radio galaxies and active galactic
nuclei~\cite{biermann}, but they are still not identified and it is
still unknown how many of them contribute to the observed cosmic ray
flux.

The observed spectrum covers about 11 orders of magnitude, from 1 GeV
to $10^{11}$ GeV, and is described by a power law $\propto E^{-\gamma}$
with two breaks at the ``knee'', at $\simeq 4 \times 10^{15}$ eV, and at
the ``ankle'', at $5 \times 10^{18}$ eV.
Above the knee the spectrum steepens from a power law index
$\gamma \simeq 2.7$ to  $\simeq 3.2$.
Above the ankle the spectrum flattens again to a power law
index $\gamma \simeq 2.8$. Cosmic rays with
energies above the ankle cannot be confined by the Galactic magnetic field,
and the lack of counterparts in our Galaxy suggests that the ankle marks a
cross-over from a Galactic component to a component of extra-galactic origin.
Data from the Fly's Eye experiment also suggest that the chemical
composition is dominated by heavy nuclei up to the ankle and by protons
beyond~\cite{Fly}.

If UHECR have an extra-galactic origin, we would expect a cutoff in
the spectrum due to the fact that in the bottom-up scenario UHECR
are assumed to be protons accelerated in powerful astrophysical
sources: Even if they can achieve, under extreme conditions, such
high energies, they will lose their energy mostly by pion production
on the microwave background.
For sources further away than a few dozen Mpc this would
predict a break in the cosmic ray flux known as
Greisen-Zatsepin-Kuzmin (GZK) cutoff~\cite{gzk}, around
$50\,$EeV. This break has not been observed by experiments such as
Fly's Eye~\cite{Fly}, Haverah Park~\cite{Haverah},
Yakutsk~\cite{Yakutsk}, Hires~\cite{Hires} and AGASA~\cite{AGASA}, which
instead show an extension beyond the expected GZK cutoff and events
above $100\,$EeV.

One of the possible solutions to the lack of observed counterparts to
the highest energy events~\cite{ssb,ES95} is to suppose the existence
of large scale intervening magnetic fields with intensity
$B\sim0.1-1\,\mu$G~\cite{ES95}, which would provide sufficient angular
deflection even for high energies and could explain the large scale
isotropy of arrival directions observed by the AGASA
experiment~\cite{AGASA} as due to diffusion.

It has been realized recently that magnetic fields as strong as
$\simeq 1 \mu G$ in sheets and filaments of large scale structures,
such as our Local Supercluster, are compatible with existing upper limits
on Faraday rotation~\cite{vallee,ryu,blasi}.

In our previous paper~\cite{isola} we considered the effects of such
strong magnetic fields in the particular case of a single
source corresponding to Centaurus A, which is a radio-galaxy located in the
southern hemisphere at a distance of 3.4 Mpc. There we employed
detailed numerical simulations for the energy spectrum and the
angular distribution of
ultra-high energy nucleons propagating in extra-galactic magnetic fields of
r.m.s. strength between 0.3 and 1 $\mu$G.
We found that this model is inconsistent with the data when
$B \simeq 0.3 \mu G$ because the angular distribution predicted is not
isotropic but concentrated around the position of the source and because
the northern hemisphere experiments should never have detected the highest
energy events for which the angular deflection is too weak to bring
the particle in the field of view of these experiments; therefore we
argued that at least a few sources within the GZK cutoff are required
to produce the observed UHECR flux.

The goal of our present paper is to elaborate more detailed constraints
on the number of sources necessary to reproduce the observed distribution,
as a function of the poorly known strength of the extra-galactic magnetic
field in our Local Supercluster. As will be explained below in more
detail, we assume a discrete distribution of sources in the
Local Supercluster permeated by magnetic fields of strength up to
$B\sim0.3 \mu\,$G.

As in our previous paper, we restrict ourselves to UHECR nucleons,
and we neglect the Galactic contribution to the
deflection of UHECR nucleons since typical proton deflection angles
in galactic magnetic fields of several $\mu$G are $\lesssim10^\circ$ above
$4\times10^{19}\, $eV~\cite{medina,anchordoqui}, and thus are small compared
to deflection in $\gtrsim0.3\mu$G fields extended over megaparsec
scales.

As statistical quantities used to test various scenarios we adopt
the angular power spectrum based on the set of spherical harmonics
coefficients $a_{lm}$, as used in Ref.~\cite{Sommers},
which is sensitive to anisotropies on large
scales, and the two-point correlation function as defined in
Ref.~\cite{Tinyakov}, which contains information on the small scale anisotropy.

As will become apparent, the statistics for these quantities is so far
limited by the small number of observed events but
the present development of large new detectors will considerably decrease
their statistical uncertainties.
In particular, the Pierre Auger experiment~\cite{auger} will
combine ground arrays measuring lateral shower cross sections
with fluorescence telescopes measuring the longitudinal shower
development. Since two of these hybrid detectors are planned,
one in the southern hemisphere currently under construction in
Argentina, and one in the northern hemisphere, full sky coverage
will be achieved, with an exposure that is practically uniform
in right ascension, and a geometrical dependence on
declination. There are furthermore plans for
space based air shower detectors such as OWL~\cite{owl} and EUSO~\cite{euso}
which may also achieve full sky coverage. For this reason it is
feasable to consider a multi-pole analysis of the angular distribution
which involves statistical estimators of integrals covering
the full sky: Since these estimators involve factors $1/ \omega_i$,
where $\omega_i$ is the exposure associated with the ith observed
direction, they are undefined if the exposure vanishes anywhere
on the sky. However, even in the absence of full-sky coverage one
can define analogous quantities and their estimators (which are
then different from the usual spherical multipoles) by simply
restricting them to the area of the sky where the exposure function
does not vanish. We will use these modified statistical quantities
to compare model predictions with the existing AGASA data.

The auto-correlation analysis provides information about the small
scale anisotropy and can be applied to partial sky coverage such
as for the AGASA experiment without restriction.
The observed data actually show significant small-scale angular clustering
(five doublets and one triplet within $2.5^\circ$ out of 57 events
above 40 EeV). This clustering has a chance probability of less than
$1\%$ in the case of an isotropic distribution. It has been pointed
out that in the presence of
turbulent extra-galactic magnetic fields of fractions of a micro
Gauss clustering could be induced by magnetic
lensing~\cite{harari,SLB99,LSB99,harari2}.
The auto-correlation analysis presented in the present paper
will demonstrate that quantitatively.

The paper is organized as follows: in section II we briefly describe
our numerical simulations, in sections III and IV we present our results on
multi-pole analysis and auto-correlation function, respectively. Section
V briefly reconsiders Centaurus A as the unique source in case of
a field as strong as a micro Gauss and in section VI we conclude.

\section{Numerical simulations}

We use the same numerical approach
used in earlier publications~\cite{isola,SLB99,LSB99}, but we
take a discrete distribution of sources centered at 20 Mpc from Earth
and distributed on a sheet with a Gaussian profile of thickness 3 Mpc
and radius 20 Mpc, with both magnetic field strength and source density
following the profile of the sheet and no sources present within
2 Mpc from the observer. This is a simple approximation to our
location in the Local Supercluster and to its shape. We also assume
that the sources inject protons with a $E^{-2.4}$ spectrum extending
up to $\simeq10^{22}\,$eV. We note that the angular distributions
are not very sensitive to assumptions on the injection spectrum.

We assume a random turbulent magnetic field with power
spectrum $\langle B(k)^2\rangle\propto k^{n_B}$ for
$2\pi/L<k<2\pi/l_c$ and $\langle B^2(k)\rangle=0$ otherwise. We use
$n_B=-11/3$, corresponding to Kolmogorov turbulence, in which case
$L$, the largest eddy size, characterizes the coherence length of the
magnetic field. For the latter we use $L\simeq1\,$Mpc, corresponding
to about one turn-around in a Hubble time. Physically one expects
$l_c\ll L$, but numerical resolution limits us to $l_c\gtrsim0.008L$. We use
$l_c\simeq0.01\,$Mpc.
The magnetic field modes are dialed on a grid in momentum
space according to this spectrum with random phases and then Fourier
transformed onto the corresponding grid in location space. The
r.m.s. strength $B$ is given by
$B^2=\int_0^\infty\,dk\,k^2\left\langle B^2(k)\right\rangle$.

Typically, 5000 trajectories are computed for each realization of
the magnetic field obtained in this way and of the source positions,
for 10-20 realizations in total. Only those trajectories
that cross a sphere of 1.75 Mpc radius around Earth (corresponding
to $5^\circ$ viewed from 20 Mpc distance) are used.
Each time such a trajectory crosses this sphere, arrival direction
and energy are registered as one event.
Each trajectory is followed for a maximal time of 10 Gyr and
as long as the distance from the observer is smaller than double
the source distance. The results do not
significantly depend on these cut-offs. Furthermore, the distance
limit is reasonable physically as it mimics a magnetic field
concentrated in the large scale structure, with much smaller
values in the voids, as generally expected.
Similar codes have been developped in Refs.~\cite{propa}.

When dialing simulated data sets from the simulated sky distributions,
one has to take into account the non-uniform exposure of the
particular experiment considered. This can be done by dialing
from the simulated distribution multiplied by an exposure
function depending on the sky solid angle $\Omega$.
This function, measured in units of km$^2$years, gives the effective
time-integrated collective area of the detector in a given
direction $\Omega$. A detector which operates continuously will have
an exposure function roughly independent of right ascension
and thus will only depend on the declination angle $\delta$. We
will only need the exposure function up to an irrelevant overall
normalization. For a detector at a single site we use the following
parameterization:
\be
  \omega(\delta) \propto \cos a_0\cos\delta\sin\alpha_m+
  \alpha_m\sin a_0\sin\delta\,,\label{exposure}
\ee
where $a_0$ is the latitude of the detector and $\alpha_m$ is zero for
$\xi > 1$, $\pi$ for $\xi < -1$, and $\cos^{-1}(\xi)$ otherwise,
where $\xi\equiv(\cos\theta_m-\sin a_0\sin\delta)/[\cos a_0\cos\delta]$.
The angle $\theta_m$ is the maximal zenith angle out to which the
detector is fully efficient ($60^{\circ}$ for Auger, $45^{\circ}$ for AGASA).
The exposure function which results for the AGASA experiment
and that we will use in the following, has been discussed,
for example, in Ref.~\cite{rr}, see in particular Fig.~2 there.

\section{The angular power spectrum}

The angular power spectrum is defined as the average $a_{lm}^2$:
\be
C(l)={1 \over 2l+1} \sum_{m=-l}^l a_{lm}^2\,,
\label{eq:1}
\ee
and the statistical estimator for the spherical harmonic coefficients
$a_{lm}$ is given by~\cite{Sommers}
\be
a_{lm}={1 \over {\cal N}}\sum_{i=1}^N {1 \over \omega_i} Y_{lm}(u^i)\,,
\label{eq:2}
\ee
where $N$ is the number of discrete arrival directions, either
of the real data, or randomly sampled from the simulated sky distributions.
Furthermore, $\omega_i$ the total experimental exposure
at arrival direction  $u^i$, ${\cal N}=\sum_{i=1}^{N}1/\omega_i$
the sum of the weights $1/\omega_i$, and
$Y_{lm}$ is the real-valued spherical harmonics function.

In order to obtain the statistical distribution of the $C(l)$
predicted by specific simulated scenarios,
we dial $C(l)$ typically $10^4$ times from the simulated distributions
[multiplied by the exposure function $\omega(\delta)$]
for each realization of the magnetic field and the source positions.

For each $l$ we plot the average over all trials and realizations
as well as two error bars. The smaller error bar (shown to the left of
the average) is the statistical error, i.e. the fluctuations due to the
finite number $N$ of observed events, averaged over all realizations,
while the larger error bar (shown to the right of the average) is the
``total error'', i.e. the statistical error plus the
cosmic variance, in other words, the fluctuations due to finite
number of events and the variation between different realizations
of the magnetic field and source positions.

To estimate the true power spectrum from Eqs.~(\ref{eq:1}) and~(\ref{eq:2})
requires data with full sky coverage and therefore at least two
detector sites such as forseen for the Auger experiment.
For its exposure function we add Eq.~\ref{exposure} for two sites
located at $a_0=-35^{\circ}$ and at $a_0=39^{\circ}$.
The AGASA experiment only has partial sky coverage and, consequently,
the true multi-pole spectrum cannot be computed from its data.
For this case we consider the quantities defined by restricting
Eq.~(\ref{eq:2}) to the sky area where $\omega(\delta)>0$.
This method is also used in the analysis of cosmic microwave
background fluctuations where window functions are used
which are unity in the observed region and zero elsewhere.
In our case this corresponds to using the AGASA exposure function
for the $\omega_i$ in Eq.~(\ref{eq:2}) for the coefficients
$a_{lm}$. This defines the modified angular power spectrum $C(l)$
both for the simulated data sets and the real data.

\begin{figure}[ht]
\includegraphics[width=0.54\textwidth,clip=true]{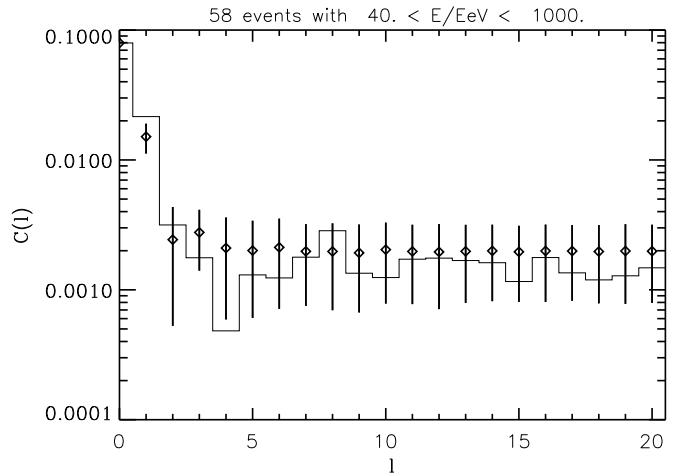}
\caption[...]{Comparison of the angular power spectrum $C(l)$,
Eqs.~(\ref{eq:1}), (\ref{eq:2}), resulting from the $N=58$
events above 40 EeV observed by AGASA (histogram), with the one
predicted for an isotropic distribution (diamonds with error bars
representing the statistical error), as a function of multipole
$l$.}
\label{isot1}
\end{figure}

We start by comparing in Fig.~\ref{isot1} the power spectra predicted
by the completely isotropic distribution with the AGASA exposure
function with the actual AGASA results which appear completely
consistent with isotropy on large scales.
Note that the increasing power for $l=0$ and $l=1$ is due
to the incomplete sky coverage of AGASA.

In the following figures, in case of full sky coverage, we show as
solid line the analytical prediction for an isotropic distribution.
In this case the power is the same for all $l$-values and decreases as
$1/N$ as the number of arrival direction increases. AGASA data
are shown as histograms.
A pure mono-pole intensity distribution is equivalent to isotropy while
the strength of other multi-poles relative to the mono-pole is a measure
of anisotropy.

Since the typical experimental angular resolution is $\simeq3^{\circ}$,
in principle information is contained in modes up to $l \sim
60$. In the following we show the values of $C(l)$ with $l$ only
up to 10 because the structure on small scales corresponding to
larger $l$ is better described by the
auto-correlation function discussed in the next section.

\begin{figure}[ht]
\includegraphics[width=0.53\textwidth,clip=true]{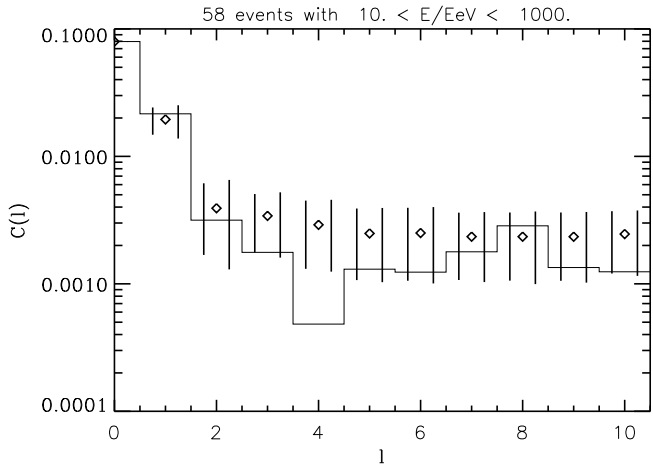}
\caption[...]{The angular power spectrum $C(l)$, Eqs.~(\ref{eq:1}),
(\ref{eq:2}), as a function of multipole $l$, obtained for the AGASA
exposure function, for $N=58$ events observed above 40 EeV, sampled
from 10 simulated realizations for $B=0.05 \mu$G
with 100 sources in the Local Supercluster. The diamonds indicate
the realization average, and the left and right error
bars represent the statistical and total (including cosmic variance
due to different realizations)
error, respectively, see text for explanations. The histogram
represents the AGASA data.}
\label{F1}
\end{figure}

\begin{figure}[ht]
\includegraphics[width=0.53\textwidth,clip=true]{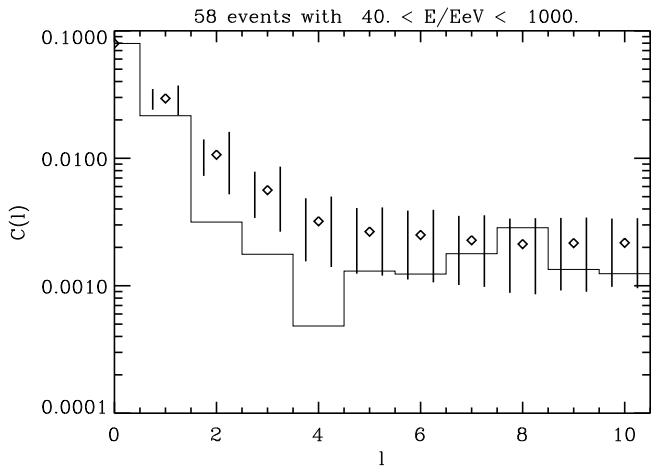}
\caption[...]{Same as Fig.~\ref{F1}, but for 400 sources.}
\label{F2}
\end{figure}

In Figs.~\ref{F1} and~\ref{F2}, we compare the angular power spectrum $C(l)$
predicted for the AGASA experiment at energies $E \geq 40 EeV$, for
magnetic field strength $B=0.05 \mu $G, and 100 and 400 sources,
respectively, with the actual AGASA data.
Both plots have been obtained for $N=58$, the present number of
events with energies above $4\times 10^{19}\,$eV observed by AGASA.

\begin{figure}[ht]
\includegraphics[width=0.50\textwidth,clip=true]{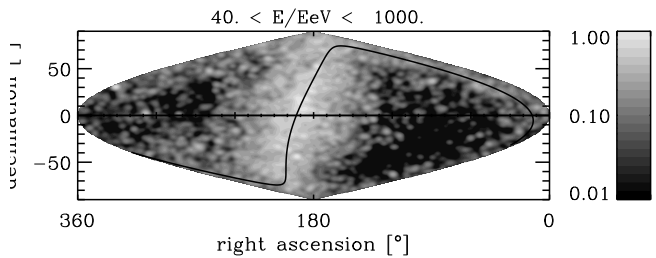}
\caption[...]{The angular image in terrestrial coordinates,
averaged over all 10 magnetic field realizations of 5000 trajectories each,
for events above 40 EeV, as seen by a detector covering all Earth with
$B=0.05\mu\,$G and 400 sources. The grey scale represents the integral
flux per solid angle. The solid line marks the supergalactic plane.
The pixel size is $1^{\circ}$ and the image has been convolved to an
angular resolution of $2.4^{\circ}$  corresponding to AGASA.}
\label{F3}
\end{figure}

The case of 100 sources seems roughly consistent with the
experimental data while the case of 400 sources shows some deviations
for the lowest multi-poles. This can be interpreted as the magnetic
field being too weak to sufficiently isotropize the arrival directions
with respect to the sources which were assumed to follow the
Local Supercluster: For a number of sources much larger than the
number of observed events $N$, it is likely that each observed event
has been produced by a different source. The number of
contributing sources is thus maximal and the fluctuations around
the assumed (non-isotropic) distribution is minimal, making the
anisotropy more visible. 
This is illustrated by Fig.~\ref{F3}, which shows
the UHECR angular distribution as seen on Earth
in terrestrial coordinates for $E \geq 40 EeV$, $B=0.05 \mu G$,
and 400 sources. 
The distribution is concentrated around the solid line
which represents the Supergalactic plane.

In contrast,
the scenario with 100 sources seems to be more sensitive to
the limited statistics due to the relatively small number of events
observed by AGASA. In fact, the statistical errors due to the small
number of events at low multi-poles is higher in Fig.~\ref{F1} than
in Fig.~\ref{F2}. 

\begin{figure}[ht]
\includegraphics[width=0.53\textwidth,clip=true]{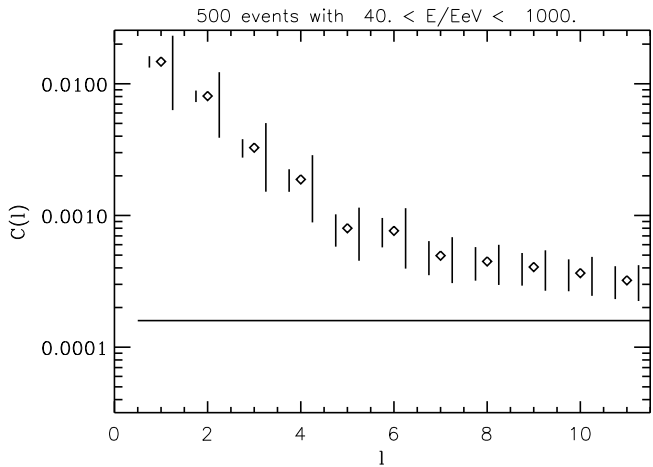}
\caption[...]{The angular power spectrum $C(l)$, Eqs.~(\ref{eq:1}),
(\ref{eq:2}), as a function of multipole
$l$, obtained for the Auger exposure function, assuming $N=500$ events
observed above 40 EeV, sampled from 10 simulated realizations
for $B=0.05 \mu$G
with 100 sources in the Local Supercluster. Average and error bars
are as in Fig.~\ref{F1}. The solid line represents the analytical
prediction for an isotropic distribution.}
\label{F4}
\end{figure}

For full sky exposure function corresponding
to the Auger parameters, and assuming 500 oberved events, we obtain
the situation shown in Fig.~\ref{F4}.
In this case the deviation from isotropy, plotted as the solid line, is
much more evident and would be easily determined by future observations.
As we will show and explain in the next section, the scenario
with 100 sources is, however, ruled out by the AGASA data from the
analysis of the auto-correlation function. For the relatively
small deflection induced by $B=0.05\mu\,$G, the number of sources
must be at least as large as the number of events observed in
different directions; much fewer than 100 sources are therefore
ruled out in this case.

\begin{figure}[ht]
\includegraphics[width=0.53\textwidth,clip=true]{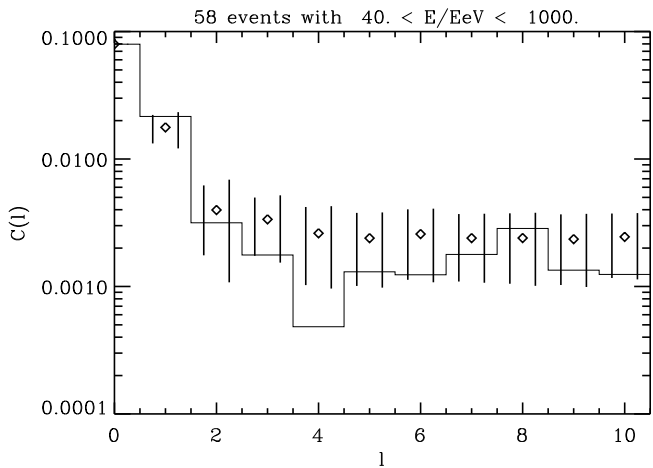}
\caption[...]{Same as Fig.~\ref{F1}, but for $B=0.3\mu\,$G and 10 sources.}
\label{F5}
\end{figure}

We now investigate whether stronger magnetic fields, by providing
larger angular deflection, might provide a better match to isotropy.
In particular, we focus on the case where $B=0.3\,\mu$ G.
In Fig.~\ref{F5} we show results for $B=0.3\,\mu $G
and 10 sources, all other assumptions being the same as in Fig.~\ref{F1}.

\begin{figure}[ht]
\includegraphics[width=0.53\textwidth,clip=true]{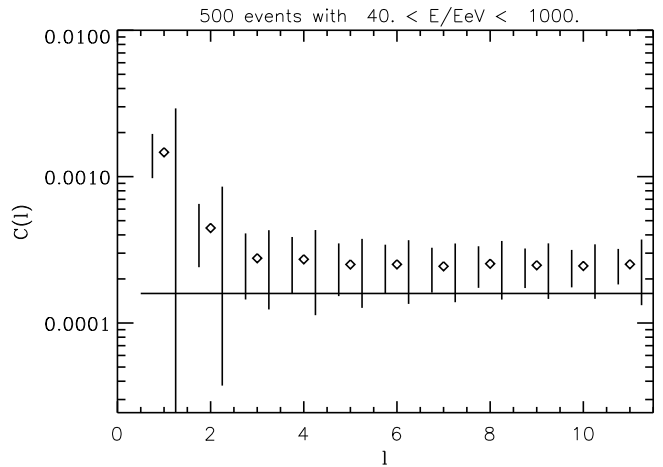}
\caption[...]{Same as Fig.~\ref{F4}, but for $B=0.3\mu\,$G, 5
sources, and 19 realizations.}
\label{F6}
\end{figure}

\begin{figure}[ht]
\includegraphics[width=0.53\textwidth,clip=true]{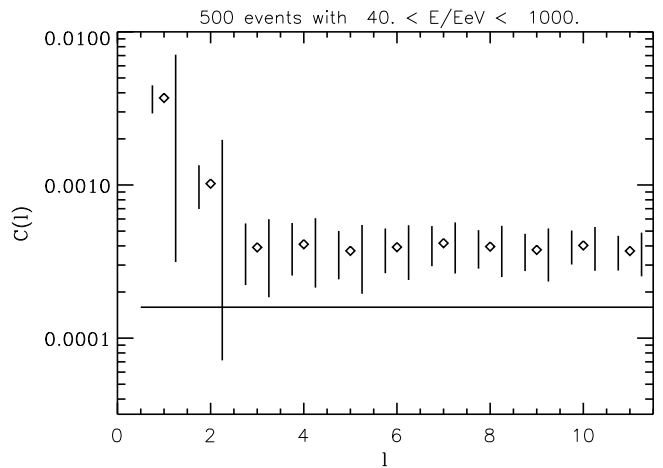}
\caption[...]{Same as Fig.~\ref{F4}, but for $B=0.3\mu\,$G and 10 sources.}
\label{F7}
\end{figure}

The distribution seems to be roughly consistent with
the data, but we also found that 5 and 100 sources result in
almost the same distribution. Since in this case the limited statistics
does not allow us to discriminate between widely different number
of sources, we turn to the case of Auger exposure with full sky coverage,
assuming 500 events observed above 40 EeV. The results are shown in
Fig.~\ref{F6} and Fig.~\ref{F7} for the case for 5 and 10
sources, respectively. The case of 100 sources is already ruled
by the auto-correlation function of the AGASA data, as will be shown
in the next section. 

Note that as for the weak field case the scenarios
shown in Fig.~\ref{F6} and~\ref{F7} predict an anisotropy that
should be easily detectable by the Pierre Auger experiment, in
contrast to the AGASA experiment (compare Fig.~\ref{F5}).
More generally, we note that the scenario with $B=0.05 \mu G$ gives
a distribution very different from the one with $B=0.3 \mu G$. Thus
future experiments with full sky coverage should be able to give
important information on the strength of the magnetic field
in the Local Supercluster by performing a multi-pole analysis.
Furthermore, in the present case of $B=0.3\mu\,$G Figs.~\ref{F6}
and~\ref{F7} show that multipoles $l\gtrsim3$ hardly depend on
the number of sources, whereas the lowest multipoles , which are
less influenced by deflection, have a noticeable dependence on
the number of sources, similarly to the case of weak field
$B\lesssim0.05\mu\,$G.

If a nearly isotropic angular distribution is confirmed by future
observations, we can conclude that for magnetic fields $B\simeq0.3\mu\,$G
a number of sources of 5-10 would be favored by the data.
As will be shown in the next section, the auto-correlation function
does not allow a much higher number of sources because
magnetic lensing would not produce sufficient clustering
on small scales. On the other hand, much fewer than 5 sources
are ruled out by the arguments given in our previous paper~\cite{isola}.

Assuming an energy independent exposure function and
using a simple $E^{-2}$ spectrum, it is possible to estimate the number
of events which will be observed in the future. In the case of Auger
observatories, for a total acceptance of $\simeq7000\,$km$^2\,$sr 
per array, in five years we should observe $\sim2200$ events above
$4 \times 10^{19}$ eV~\cite{Sommers}. Here we have been conservative
and used $N=500$.

\section{Auto-correlation function}

For the auto-correlation function we follow the same approach used in
Ref.~\cite{Tinyakov}. We start from either actual data or from
a randomly generated set of $N$ events
dialed from the simulated distributions, multiplied by the exposure
function. For each event we divide the sphere into concentric bins
with a fixed angular size $\Delta\theta$, and we count the number of
events falling into each bin. We then divide by twice the solid
angle size $S(\theta)$ of the corresponding bin, arriving at
\be
N(\theta)=\frac{1}{2S(\theta)}\sum_{j \neq i}R_{ij}(\theta)\,,
\label{auto}
\ee
where
$$R_{ij}(\theta)=\left\{\begin{array}{ll}
1 & \mbox {if $\theta_{ij}$ is in same bin as $\theta$}\\
0 & \mbox{otherwise}
\end{array}\right.\,.$$
We note that the auto-correlation function in the strict sense
would include a factor $({\cal N}^2\omega_i\omega_j)$ under the
sum in Eq.~(\ref{auto}). However, the differences are small and in
any case we are free to choose any statistical quantity as long as
the same quantity and its fluctuations are used to compare
simulations and data.

In analogy to the previous section, for each magnetic field and
source position realization we dial $N(\theta)$ $10^4$ times from
the simulated distributions in order to obtain its average and
variances for which we plot the same two error bars as for the power
spectrum. The histograms shown subsequently represent again the
result for the AGASA data, where the sharp peak at small separation
angles results from the six observed clusters. We have verified
that using incorrect exposure functions in general destroys the
agreement found between a simulated isotropic distribution and the
data at large $\theta$~\cite{AGASA}. The observed distribution of
events thus reflects the non-uniform exposure~\cite{rr,Tkachev}.

\begin{figure}[ht]
\includegraphics[width=0.53\textwidth,clip=true]{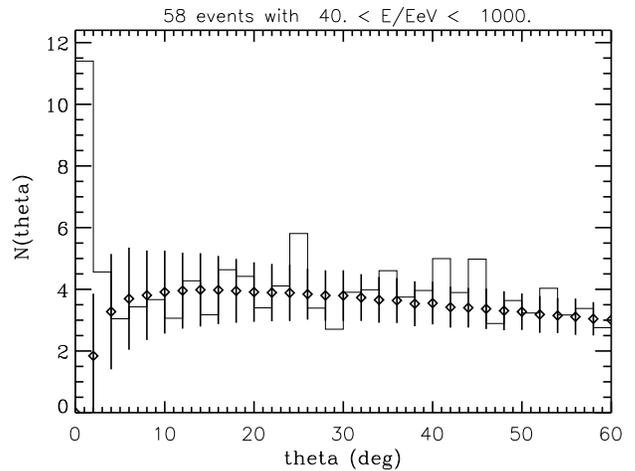}
\caption[...]{Comparison of the angular correlation function $N(\theta)$,
Eq.~(\ref{auto}), resulting from the $N=58$
events above 40 EeV observed by AGASA (histogram), with the one
predicted for an isotropic distribution (diamonds with error bars
representing the statistical error), as a function of angular
distance $\theta$. A bin size $\Delta\theta=2^\circ$ was used.}
\label{isot2}
\end{figure}

We start  by comparing the autocorrelation function for the
real AGASA data with the isotropic distribution in Fig.~\ref{isot2}.
This demonstrates that the AGASA data are completely consistent
with isotropy except at scales larger than a few degrees.

\begin{figure}[ht]
\includegraphics[width=0.53\textwidth,clip=true]{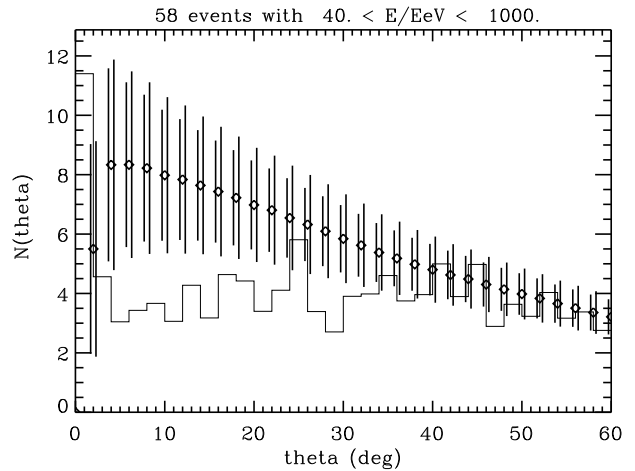}
\caption[...]{The angular correlation function $N(\theta)$,
Eq.~(\ref{auto}), as a
function of angular distance $\theta$, obtained for the AGASA exposure
function, for $N=58$ events observed above 40 EeV, sampled from
10 simulated realizations for $B=0.05 \mu$G with 100 sources in the Local
Supercluster.
Average and error bars are as in Fig.~\ref{F1}. The histogram again
represents the AGASA data. A bin size $\Delta\theta=2^\circ$ was used.}
\label{F8}
\end{figure}

In Fig.~\ref{F8} we show the angular correlation function for $N=58$ events
with energies $E\geq 40 EeV$, predicted by simulations
with $B=0.05 \mu$G and 100  sources, using the AGASA exposure function.

This case shows no correlation at angles as small as the angular
resolution, where AGASA shows a peak, whereas there are strong
correlations at larger angles, which is not
consistent with the observed isotropic distribution at large scales.
This also corresponds to the fact that in the case of weak magnetic fields
we expect that clusters just reflect the point-like sources but
if the number of sources is much larger than the number of observed
events $N$, each source constributes at most one events and clustering
is not possible. As remarked in the previous section, a much smaller
number of sources is not possible either due to the large number
of observed arrival directions.

\begin{figure}[ht]
\includegraphics[width=0.53\textwidth,clip=true]{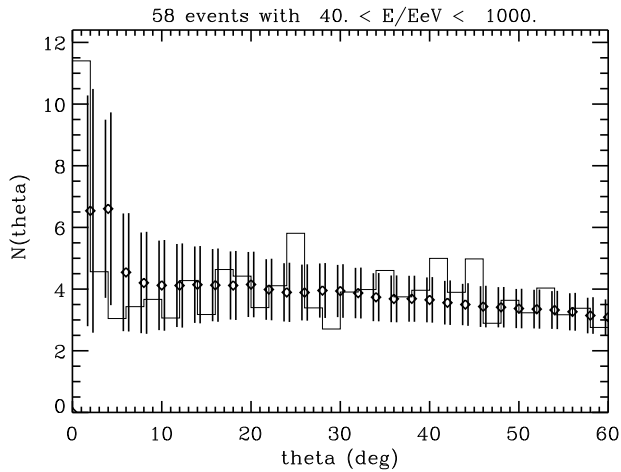}
\caption[...]{Same as Fig.~\ref{F8}, but for $B=0.3\mu\,$G and 5
sources with 20 realizations.}
\label{F9}
\end{figure}

\begin{figure}[ht]
\includegraphics[width=0.53\textwidth,clip=true]{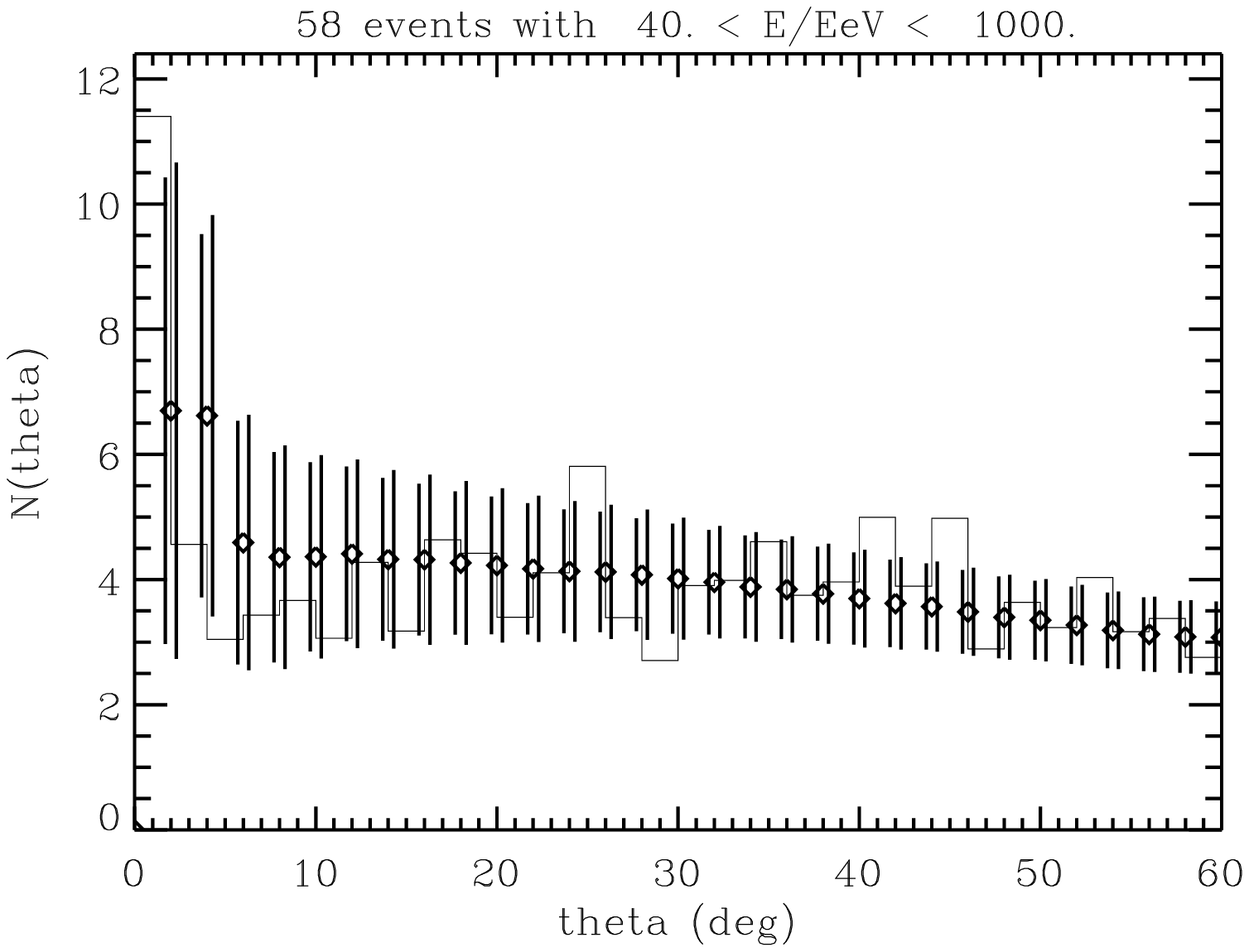}
\caption[...]{Same as Fig.~\ref{F8}, but for $B=0.3\mu\,$G and 10 sources.}
\label{F10}
\end{figure}

\begin{figure}[ht]
\includegraphics[width=0.53\textwidth,clip=true]{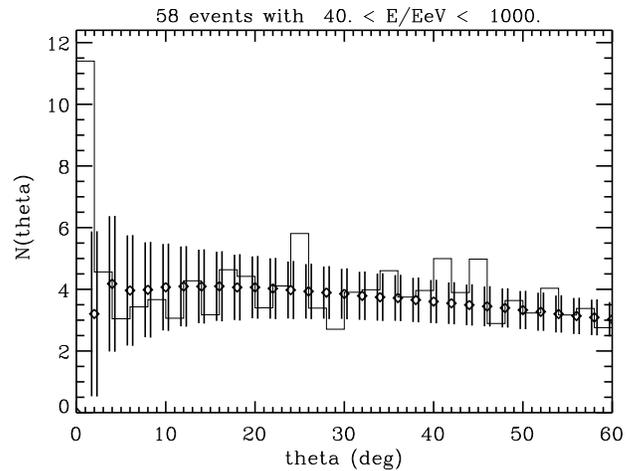}
\caption[...]{Same as Fig.~\ref{F8}, but for $B=0.3\mu\,$G and 100
sources with 18 realizations.}
\label{F11}
\end{figure}

In Figs.~\ref{F9},~\ref{F10}, and~\ref{F11} we show the angular
correlation functions predicted by scenarios with
$B=0.3 \mu$G, with 5 and 10 and 100 sources, respectively.
In the case of 100 sources the simulated distribution do not show any
correlation at small scales. Similarly to the weak field case,
this can be understood due to the fact that the source images
produced by magnetic lensing contain at most one event if the
number if sources is much larger than $N$. Note also that cosmic
variance becomes very small for $N\gtrsim100$.

\begin{figure}[ht]
\includegraphics[width=0.53\textwidth,clip=true]{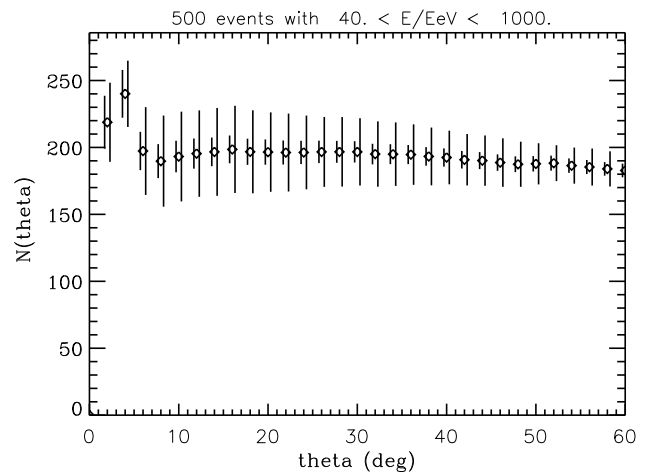}
\caption[...]{Same as Fig.~\ref{F10}, but obtained for
the Auger exposure function, assuming $N=500$ events observed
above 40 EeV.}
\label{F12}
\end{figure}

We obtain the same result for $B=0.1 \mu$G and 100 sources.
Thus we can argue that 100 is an approximate current upper
limit for the number of sources. On the other hand for 5-10
sources the simulated auto-correlation function seems to be in
agreement with the observed clustering at small scales. In Fig.~\ref{F12}
we show for one of these cases what could be expected for
Auger exposures.
This demonstrates that, for the amount of data expected with next 
generation experiments, the statistics will be dominated by cosmic
variance instead of the limited number of events observed, as is
presently the case.

\section{Only one source: Centaurus A}
\begin{figure}[ht]
\includegraphics[width=0.53\textwidth,clip=true]{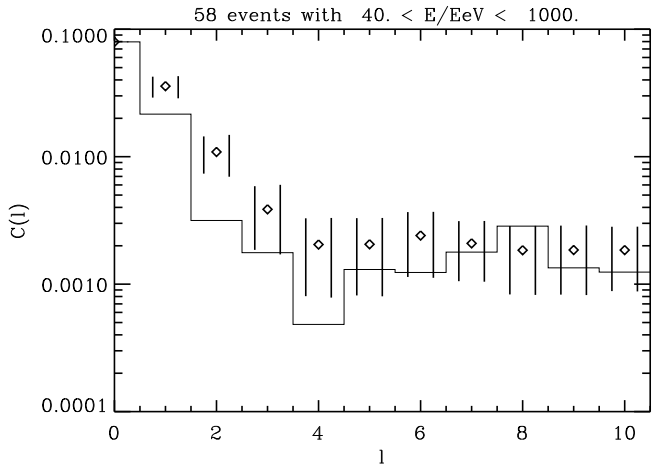}
\caption[...]{Same as Fig.~\ref{F1}, but for the scenario with
Centaurus A at 3.4 Mpc distance as the single source, a field
of $1\mu\,$G, and 20 realizations.}
\label{F13}
\end{figure}

\begin{figure}[ht]
\includegraphics[width=0.53\textwidth,clip=true]{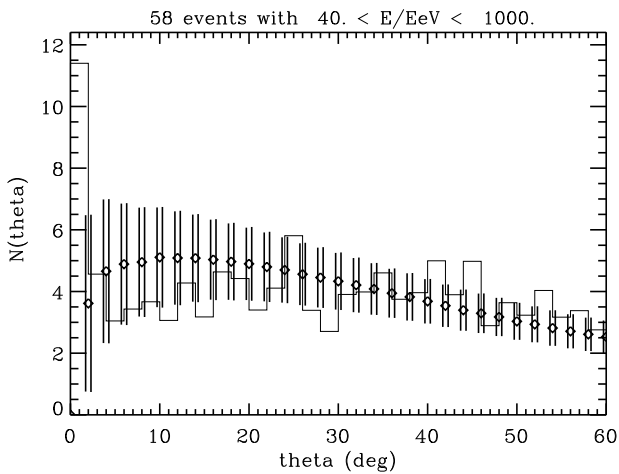}
\caption[...]{Same as Fig.~\ref{F8}, but for the scenario with
Centaurus A at 3.4 Mpc distance as the single source, a field
of $1\mu\,$G, and 20 realizations.}
\label{F14}
\end{figure}

Now we briefly reconsider the model discussed in our previous
paper~\cite{isola}, with Centaurus A as single source,
and $B = 1\mu\,$G. The predictions for the angular power spectrum
and the auto-correlation function are compared with the AGASA
data in Figs.~\ref{F13} and~\ref{F14}, respectively.

\begin{figure}[ht]
\includegraphics[width=0.53\textwidth,clip=true]{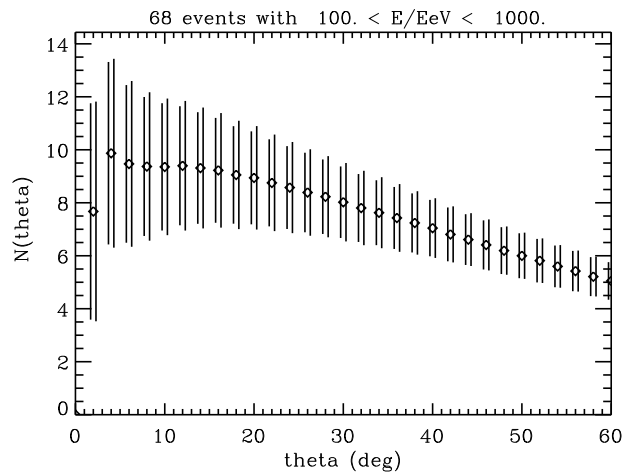}
\caption[...]{The angular correlation function $N(\theta)$ as a
function of angular distance $\theta$, obtained for the Auger exposure
function, assuming $N=68$ events observed above 100 EeV, for the
scenario with Centaurus A at 3.4 Mpc distance as the single source,
a field of $1\mu\,$G, and 20 realizations. The bin size is again
$\Delta\theta=2^\circ$.}
\label{F15}
\end{figure}

The angular power spectrum shows a $\simeq 3\sigma$
deviation from the data at $l=2$. Furthermore,
the auto-correlation function does not show significant correlations
at angular resolution scales, in contradiction to the data.
This is due to the fact that for a magnetic field as strong as 1$\mu$G,
we are in a range of energies where many overlapping images the
source are produced~\cite{harari2}. Correlations up to relatively
large scales would only appear at energies above $\simeq10^{20}\,$eV,
as can be seen in Fig.~\ref{F15} which was produced for the
Auger exposure function.

\section{Discussion and Conclusions}

In the present work we assumed a discrete distribution of sources distributed
in the Local Supercluster and estimated the number of sources
necessary to reproduce the experimental data, in dependence on the
typical strength of the extra-galactic magnetic fields permeating
the Local Supercluster. As statistical quantities for this analysis
we used spherical multi-poles and the auto-correlation function.
We found that for weak magnetic fields $\lesssim0.05\mu\,$G the
simulation predictions appear to be not consistent with the observed
distribution for any number of sources because the magnetic field is
too weak to isotropize the anisotropic distribution associated with the
Supergalactic plane. Full sky experiments of the size of the Pierre Auger
project will be sensitive to the difference in the distribution of multi-poles
between weak and strong magnetic fields, which thereby could give
direct informations about the strength of the magnetic fields.
For stronger magnetic fields $\simeq0.3\mu\,$G we found that the
number of sources is constrained. In our previous paper~\cite{isola}
we already showed that a single source cannot reproduce the observed
isotropic distribution. Here we found that for $\gtrsim100$ sources the
auto-correlation function does not reproduce the correlations observed
at small scales. This can be interpreted by the fact that for a number of
sources much higher than the number of observed events there are more
source images produced by lensing than observed events and thus
clustering is not observable. Therefore, the current upper limit
on the number of contributing sources is $\simeq100$ in this case.
On the other hand, a number of sources around 10 seems to reproduce
quite well the observed small scale clustering. We also showed that
the model with Centaurus A as the only source and very strong fields
$\simeq1\mu G$, considered as marginally consistent in our previous
paper~\cite{isola}, is not consistent with isotropy
at large scales due to a predicted quadrupole
deviation from isotropy and because the auto-correlation function
is not consistent with the clustering at small scales observed by AGASA.
We conclude then that a distribution of $\simeq10$ sources in the
Local Supercluster, with magnetic fields in the sub micro Gauss
range, could reproduce at least current observations. Our approach
can equally be applied to other source distributions.

Due to the still sparse statistics of current data, in the present paper
we refrained from quantifying the statistical significance of
deviations between models and data, because small number fluctuations
are in general not Gaussian and, for different multipoles and
separation angles, can be correlated. Results based on comparison
with the AGASA data presented here should rather be understood as
suggestive tendencies. Quantitative significances can be obtained
by determining
in how many simulated trials a certain quantity, such as the multi-poles
and auto-correlations studied here, or certain combinations
thereof, show deviations from the data of opposite sign to the
average deviation. We leave that to a study with data of much
higher statistics above $\simeq10^{19}\,$eV than available today,
as expected from future full-sky experiments. These experiments
will put much more stringent constraints both on the number of sources and
the magnetic field strength.

\section*{Acknowledgements} We thank Martin Lemoine for
past and ongoing collaborations related to the development
of numerical propagation codes. CI also wishes to thank 
Olivier Dor{\'e} and Lorenzo Sorbo for useful suggestions on the manuscript.



\begin{thebibliography}{9}

\bibitem{reviews} for recent reviews see J.~W.~Cronin, Rev.~Mod.~Phys.
71 (1999) S165; P.~Bhattacharjee and G.~Sigl, Phys.~Rept. 327 (2000) 109;
A.~V.~Olinto, Phys.~Rept. 333-334 (2000) 329; X.~Bertou,
M.~Boratav, and A.~Letessier-Selvon, Int.~J.~Mod.~Phys. A15 (2000) 2181.

\bibitem {biermann} see, e.g., P.~L.~Biermann, J.~Phys.~G: Nucl.~Part.~Phys.
23 (1997) 1.

\bibitem{Fly} D. J. Bird et al., Phys. Rev. Lett.71 (1993) 3401~;
  Astrophys. J. 424 (1994) 491~; ibid.441 (1995) 144.

\bibitem{gzk} K.~Greisen, Phys.~Rev.~Lett. 16 (1966)
748; G.~T.~Zatsepin and V.~A.~Kuzmin, Pis'ma
Zh. Eksp. Teor. Fiz. 4 (1966) 114 [JETP. Lett. 4 (1966) 78].

\bibitem{Haverah} See, e.g., M. A. Lawrence, R. J. O. Reid, and
  A. A. Watson, J.Phys. G 17 (1991) 733, and references
  therein~; see also {\sf http~://ast.leeds.ac.uk/haverah/hav-home.html}.

\bibitem{Yakutsk} N. N. Efimov et al., Proc. International Symposium on
  {\it Astrophysical Aspects of the Most Energetic Cosmic
  Rays}, eds.  M. Nagano and F. Takahara (Worls Scientific Singapore,
  1991) p.20~; B. N. Afnasiev, Proc. of International Symposium on {\it
  Extremely High Energy Cosmic Rays~: Astrophysics and Future
  Observatoires}, ed. M. Nagano (Instiute for Cosmic Ray Research,
  Tokyo, 1996), p.32.

\bibitem{Hires} D. Kieda et al., Proc. of the 26th ICRC, Salt Lake,
  1999~; see also {\sf http~://www.physics.utah.edu/Resrch.html}.

\bibitem{AGASA} Takeda et al., Astrophys. J. 522 (1999) 225;
M.Takeda et al., Phys. Rev. Lett. 81 (1998) 1163; Hayashida et al.,
e-print astro-ph/0008102; see also
{\sf http~://www-akeno.icrr.u-tokyo.ac.jp/AGASA/}.

\bibitem{ssb} G.~Sigl, D.~N.~Schramm, and P.~Bhattacharjee,
Astropart.~Phys. 2 (1994) 401.

\bibitem{ES95} J.~W.~Elbert, and P.~Sommers, Astrophys.~J. 441
(1995) 151;

\bibitem{vallee} J.~P.~Vall{\'e}e, Fundamentals of Cosmic
Physics, Vol.~19 (1997) 1.

\bibitem{ryu}D.~Ryu, H.~Kang, and P.~L.~Biermann,
Astron.~Astrophys. 335 (1998) 19.

\bibitem{blasi} P.~Blasi, S.~Burles, and A.~V.~Olinto, Astrophys.~J.
514 (1999) L79.

\bibitem{isola} C. Isola, M. Lemoine, and G. Sigl, Phys.~Rev.~D65
(2002) 023004.

\bibitem{medina} G.~Medina-Tanco, E.~M.~De~Gouveia~Dal~Pino,
and J.~E.~Horvath, e-print astro-ph/9707041.

\bibitem{anchordoqui} L.~A.~Anchordoqui and H.~Goldberg,
e-print hep-ph/0106217; T.~Stanev, Astrophys.~J. 479 (1997) 290.

\bibitem{Sommers} P. Sommers, Astropart.~Phys. 14 (2001) 271.

\bibitem{Tinyakov} P.G. Tinyakov and I.I. Tkachev JETP Lett. 74 (2001) 1; M. Takeda et all. Proceedings of ICRC 2001:345.

\bibitem{auger} J.~W.~Cronin, Nucl.~Phys.~B (Proc.~Suppl.) 28B
(1992) 213; The Pierre Auger Observatory Design Report (2nd
edition), March 1997; see also
{\sf http://http://www.auger.org/} and
{\sf http://www-lpnhep.in2p3.fr/auger/welcome.html}.

\bibitem{owl} D.~B.~Cline, F.~W.~Stecker, OWL/AirWatch science
white paper, e-print astro-ph/0003459;
see also {\sf http://lheawww.gsfc.nasa.gov/docs/gamcosray/hecr/OWL/}.

\bibitem{euso} R.~Benson, J.~Linsley,
Southwest.~Reg.~Conf.~Astron.~\&~Astrophys. 7 (1992) 161;
see also {\sf http://www.ifcai.pa.cnr.it/Ifcai/euso.html}.

\bibitem{harari} D.~Harari, S.~Mollerach, and E.~Roulet,
JHEP 08 (1999) 022; ibid. 0002 (2000) 035;  ibid. 0010 (2000) 047.

\bibitem{SLB99} G.~Sigl, M.~Lemoine, and P.~Biermann, Astropart.~Phys.
10 (1999) 141.

\bibitem{LSB99} M.~Lemoine, G.~Sigl, P.~Biermann, e-print astro-ph/9903124.

\bibitem{harari2} D. Harari, S. Mollerach, E. Roulet and F. Sanchez,
e-print astro-ph/0202362.

\bibitem{propa} G.~Medina Tanco, e-print astro-ph/9808073;
A.~Achterberg, Y.~A.~Gallant, C.~A.~Norman, and D.~B.~Melrose,
e-print astro-ph/9907060;
T.~Stanev et al., Phys.~Rev.~D62 (2000) 093005;
Y.~Ide, S.~Nagataki, and S.~Tsubaki, e-print astro-ph/0106182.

\bibitem{rr} S.~Razzaque and J.~P.~Ralston, e-print astro-ph/0110045.

\bibitem{Tkachev} P.~Tinyakov and I.~Tkachev, private communication.

\end{thebibliography}
\end{document}